\documentclass[slac_one]{revtex4}
\usepackage{graphicx}
\usepackage{fancyhdr}
\begin{document}

\title{Source of CP Violation for the Baryon Asymmetry of the Universe
} 

%

\author{George W.S. Hou$^{1,2,3,4}$}
\affiliation{$^{1}$Department of Physics,
 $^{2}$Institute of Astrophysics,
 $^{3}$National Center for Theoretical Sciences,
 $^{4}$
      Leung Center for Cosmology and Particle Astrophysics,
       National Taiwan University,
       Taipei, Taiwan 10617, R.O.C.
 }
%

\begin{abstract}
How to account for the matter predominance of our Universe is a
fundamental issue at the core of our existence. One condition is
$CP$ violation, but the Standard Model falls short by more than
$10^{-10}$. Taking cue from a recent result from the $B$
factories, we find that a fourth quark generation can provide
enhancement by a factor of $10^{13}$ or more. This could be the
source of $CP$ violation for the baryon asymmetry of the Universe.
The main source of enhancement is the large Yukawa couplings of
the heavy $t^\prime$ and $b^\prime$ quarks. With indications for a
new, large $CP$ violating phase $\sin2\Phi_{B_s}$ emerging at the
Tevatron, our suggestion can be verified or refuted at the LHC in
the next few years.
\end{abstract}

\maketitle

\thispagestyle{fancy}


\section{THE LORE THAT DESPAIRS THE EXPERIMENTER} 

Having crossed the boundaries between theory and experiment, I
attest that the experimenter feels like a hapless ant crawling on
a desk, as far as searching for New Physics (NP) $CP$ violation
(CPV) is concerned.

Take, for example, the Belle \textit{Nature}
paper~\cite{belleDeltaA} on the difference in direct CPV between
$B^+$ and $B^0$ ($\Delta {\cal A}_{K\pi}$). In reporting a large
deviation, Belle cited the Baryon Asymmetry of the Universe (BAU)
as the reason to pursue CPV studies, but immediately admitted that
all data support the unique Kobayashi--Maskawa (KM,~\cite{KM})
source of CPV in the Standard Model (SM), which is ``known to be
too small" (by $10^{-10}$~\cite{Peskin} at least) for BAU. Because
the gap (which the general experimenter knows only vaguely)
between SM and the heavenly BAU is so large, it appears
insurmountable, no matter what is found in the laboratory.

It is truly remarkable that the SM has~\cite{KRS85} all the
necessary ingredients for baryogenesis, i.e. the Sakharov
conditions of baryon number violation, CPV, and deviation from
equilibrium (in the very hot early Universe). The agony is the
insufficiency in the latter two: CPV is way to small, while the
electroweak phase transition (EWPhT) seems only a crossover.
We see no antibaryons in our Universe, i.e. $n_{\bar{\cal
B}}/{n_\gamma} = 0$, while ${n_{\cal B}}/{n_\gamma} = (6.1 \pm
0.2) \times 10^{-10}$ (WMAP); BAU is 100\%. But the folklore is
that SM falls short by $10^{-10}$. The source of this is the
Jarlskog invariant~\cite{Jarlskog},
\begin{eqnarray}
J &=& (m_t^2 - m_u^2)(m_t^2 -  m_c^2)(m_c^2 - m_u^2)
      (m_b^2 - m_d^2)(m_b^2 -  m_s^2)(m_s^2 - m_d^2)\, A,
 \label{eq:J3}
\end{eqnarray}
which incorporates all requirements for CPV to be nonvanishing,
where $A$ is twice the area of any unitarity triangle. Note that
$J$ has dimensions $M^{12}$. To compare with ${n_{\cal
B}}/{n_\gamma}$, one typically normalizes by the EWPhT temperature
$T \sim 100$ GeV (or roughly the v.e.v. scale). Putting in quark
masses, and our knowledge that $A \simeq 3\times 10^{-5}$, one
immediately finds $J/T^{12} \sim 10^{-20}$, which falls short by
$10^{-10}$. The main source of suppression is the smallness of
light quark masses. The situation is in general much worse, since
there are coupling constant factors as well.

\centerline{
 --- \ \emph{We observe that, by extending from 3 to 4 quark
     generations, one can enhance Eq.~\ref{eq:J3} by over $10^{13}$}
 \ ---}

The thread that lead to this observation appeared concurrent with
the 2004 observation of direct CPV in $B^0 \to K^+\pi^-$ mode,
i.e. the first hint of $\Delta {\cal A}_{K\pi} \equiv {\cal
A}_{K^+\pi^0} - {\cal A}_{K^+\pi^-} \neq 0$. Written into the
Belle paper~\cite{belle04} at that time, it was noted that if the
electroweak penguin $P_{\rm EW}$ ($Z$ penguin really) is the
source of this apparent difference, then NP CPV is implied.
However, as is well known, the so-called color-suppressed tree
diagram $C$ could also generate $\Delta {\cal A}_{K\pi} \neq 0$.
Although Peskin~\cite{Peskin} stressed the equal possibility of
$C$ vs $P_{EW}$ origins in his companion \textit{Nature} paper,
privately he is ``very skeptical that the new Belle result is new
physics".

So, with the gap of $10^{-10}$ in mind, the hapless ant crawls on.

\section{GOING UP A HILL, ... WHICH MAY BECOME A MOUNTAIN}

\subsection{Crawling Up a Hill}

Noticing that the $P_{\rm EW}$ (or the $Z$ penguin), where the $Z$
is radiated off a virtual top or $W$ in a $b\to s$ loop and turns
into a $\pi^0$ (but not a $\pi^-$), I recalled my first B paper on
$b\to s\ell^+\ell^-$~\cite{HWS87}. Naive counting would lead one
to conclude that the photonic penguin diagram, at ${\cal O}(\alpha
G_F)$, would dominate over the $Z$ penguin, at ${\cal O}(G_F^2)$.
Even if one notes that the two differ by $m^2$ in dimensions, one
would still have $G_F^2 m_b^2 \ll \alpha G_F$. But it turns out,
by direct computation, or by argument of conserved vector current
vs spontaneous electroweak symmetry breaking (EWSB), that the $Z$
penguin behaves as $G_F^2 m_t^2$ and actually dominates. This is
called \emph{nondecoupling} of heavy chiral quark masses in SM.
I therefore embarked on crawling up the little hill of adding a
4th generation.

But this usually appears as running against a wall in a quixotic
way; the 4th generation has long been viewed by many as ruled out
already, by neutrino counting and electroweak precision tests
(EWPrT). However, we now know that neutrinos have mass, which
calls for New Physics, while Kribs \textit{et al.}~\cite{KPST07}
recently pointed out that the 4th generation is not in such great
conflict with EWPrT.
In any case, we demonstrated, both at LO~\cite{HNS05} and
NLO~\cite{HLMN07} in PQCD factorization approach (the only one
that predicted the size and sign of ${\cal A}_{K^+\pi^-}$), that
the 4th generation can generate the observed $\Delta {\cal
A}_{K\pi}$. In Ref.~\cite{HLMN07} we showed that $\Delta {\cal
S}_{K^0\pi^0}$ and $\Delta {\cal S}_{\phi K_S}$ moved downwards by
$\sim -0.1$, which is the right direction and consistent with
current data. Both the sign and strength are nontrivial.

\subsection{Becoming a Mountain\,?}

$\Delta {\cal A}_{K\pi} \simeq 15\% > -{\cal A}_{K^+\pi^-} \sim
10\%$ is rather large for a NP effect. Given that the $b\bar
s\leftrightarrow \bar b s$ box diagram has similar
$m_{t^{(\prime)}}$ dependence as in the $b\to s$ $Z$ penguin, with
the CDF measurement of $B_s$ mixing, a very sizable,and {\it
negative}, mixing-dependent CPVis predicted~\cite{HNS07} for
$B_s\to J/\psi \phi$, i.e.
\begin{eqnarray}
\sin2\Phi_{B_s} \equiv -\sin2\beta_s \sim -0.5\; {\rm \ to\ }
-0.7,
 \label{eq:sin2PhiBs}
\end{eqnarray}
compared with $\sin2\Phi_{B_s}\vert^{\rm SM} \sim -0.04$. The
range of $-0.4$ to $-0.7$ was already predicted in
Ref.~\cite{HNS05} and reported~\cite{ICHEP06} at ICHEP 2006 in
Moscow. The improvement of Eq.~\ref{eq:sin2PhiBs} came with the
more precisely determined $B_s$ mixing, while the sign is
determined by the sign of $\Delta {\cal A}_{K\pi}$. So at ICHEP
2006, I already asked ``Can large CPV in $B_s$ mixing be measured
@ Tevatron\,?", and pronounced that the case is good for the
Tevatron (vs LHCb, which comes on later).

Interestingly, by end of 2007, CDF reported~\cite{phisCDFtag}
indications for $\sin2\beta_s$ that is consistent with
Eq.~\ref{eq:sin2PhiBs}, but less consistent with the SM
expectation. By this conference, the D$\emptyset$
measurement~\cite{phisDzerotag} and a CDF update~\cite{Tonelli}
both confirm this trend, and the combined deviation from SM is
now~\cite{Paulini} more than 2$\sigma$, with central value of
$\sim -0.6\,$!

This incredible development makes 2009--2010 very interesting,
whether LHCb arrives on the scene or not.

\section{SOARING TO THE STARRY HEAVENS}

Heavy SM chiral quark effects are nondecoupled in the box and $Z$
penguin diagrams. The source is both because of the subtleties of
spontaneous EWSB, and that heavy quark masses are due to large
Yukawa couplings to the v.e.v. This I knew since twenty some
years. Stimulated by large $\Delta {\cal A}_{K\pi}$, in the past 4
years I could not stop from pushing the work on the 4th
generation, utilizing large $t'$ Yukawa couplings, and CPV phase
in $V_{t's}V_{t'b}^*$.

I cannot remember when and how, but one day the ``YuReKa(wa)"
moment came: large Yukawa couplings can modify Eq.~\ref{eq:J3}\,!
If one shifts by one generation with 4th generation SM (SM4), then
Eq.~\ref{eq:J3} becomes~\cite{CPVBAU}
\begin{eqnarray}
J_{(2,3,4)}^{sb}
 &\simeq& (m_{t^\prime}^2-m_c^2)(m_{t^\prime}^2-m_t^2)(m_t^2-m_c^2)
          (m_{b^\prime}^2-m_s^2)(m_{b^\prime}^2-m_b^2)(m_b^2-m_s^2)\,
                 A_{234}^{sb}.  
 \label{eq:J234}
\end{eqnarray}
The notation will be clarified soon, but it is clear that the
difference of light quark mass pairs,
$(m_c^2-m_u^2)(m_b^2-m_d^2)(m_s^2-m_d^2)$ now all drop out, and
one gains in the mass factors (assuming $m_{b',t'} \sim 300$ GeV)
by $10^{13}$\,! For the change in CPV ``area" $A$, if the hints
from $\Delta {\cal A}_{K\pi}$ and $\sin2\Phi_{B_s}$ hold up, one
could gain a further factor of 30.

\begin{figure*}[t]
\centering
 \vskip-1.8cm
\includegraphics[width=130mm]{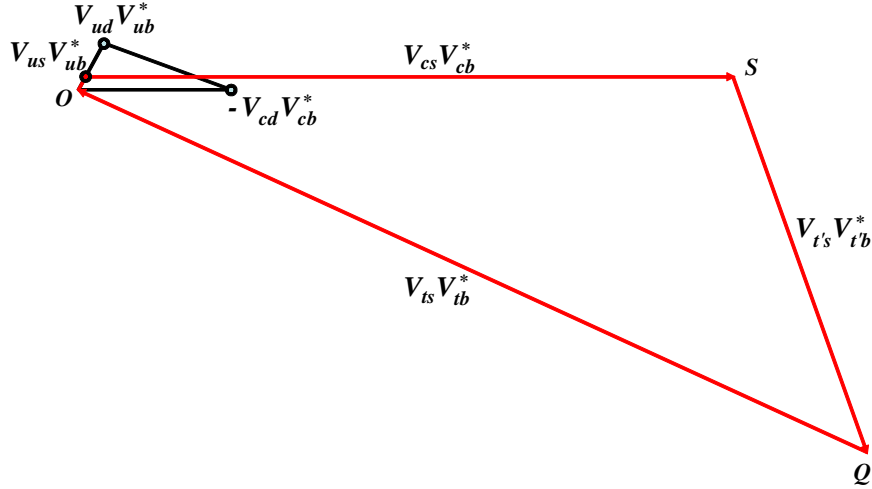}
 \vskip-1.5cm
\caption{The small SM-like $b\to d$ triangle (gives area
  $A/2$ in Eq.~\ref{eq:J3}), and SM4 $b\to s$ quadrangle
  (gives area $A_{234}^{sb}/2$ in Eq.~\ref{eq:J234}).
  The large area, and the size and orientation of phase angle
  at $S$, lead to the prediction that $\sin2\Phi_{B_s}$ is
  large and negative.
  }
 \label{bdbsTri}
\end{figure*}

To illustrate this last point, we show in Figure~\ref{bdbsTri} the
$b\to s$ quadrangle corresponding to the SM4 unitarity relation
$V_{u s}V^*_{u b} + V_{c s}V^*_{cb} + V_{t s}V^*_{t b} +
V_{t^\prime s}V^*_{t^\prime b} = 0$, together with the SM3 $b\to
d$ triangle $V_{u d}V^*_{u b} + V_{c d}V^*_{cb} + V_{t d}V^*_{t
b}= 0$. The latter is from the current 3 generation fit to all
data, the success of which lead to KM receiving the 2008 Nobel
Prize. For the former, it comes from the program~\cite{HNSprd05}
that started with $\Delta {\cal A}_{K\pi}$ (fixes
$V_{t's}V_{t'b}^*$ for given $m_{t'}$), but incorporating the
$Z\to b\bar b$ and rare kaon constraints on $V_{t'b}V_{t'b'}^*$
and $V_{t'd}V_{t's}^*$, using unitarity of $4\times 4$ CKM matrix.

We note that, if one draws the line linking $S$ and $O$ in
Figure~\ref{bdbsTri}, the rather squashed and elongated triangle
corresponds to $V_{u s}V^*_{u b} + V_{c s}V^*_{cb} + V_{t s}V^*_{t
b} = 0$ for SM3, the 3 generation SM. This triangle has the same
area $A/2$ as the $b\to d$ triangle. It is the very tiny phase
angle of the $b\to s$ triangle in SM3 at the vertex $S$ that gives
rise to the very small value of $\sin2\Phi_{B_s}\vert^{\rm SM3}$.
The sign, which is opposite to $\sin2\Phi_{B_d}\vert^{\rm SM3}
\equiv \sin2\phi_1/\beta$, is because the ``orientation" is
opposite that of the SM3 $b\to d$ triangle. The large phase angle
in SM4 at vertex $S$ leads to the large area of the quadrangle, or
${A_{234}^{sb}}/{A} \sim 30$, hence our prediction of
Eq.~\ref{eq:sin2PhiBs}.

Why do the $b\to d$ processes give a triangle, rather than a
quadrangle, if there are 4th generation effects lurking in $b\to
s$ transitions? This question was dealt with in
Ref.~\cite{HNSprd05}: with large $V_{t's}V_{t'b}^*$ (including CPV
phase), after taking into account the $Z\to b\bar b$ and rare kaon
constraints, the actual $b\to d$ quadrangle mimics the SM3
triangle, or $b\to d$ transitions are SM-like. This is a
nontrivial test, and indeed, another possible solution is rejected
by this.

\section{DISCUSSION AND CONCLUSION}

\subsection{Towards Solution of BAU}

The original Jarlskog invariant of Eq.~\ref{eq:J3} was
derived~\cite{Jarlskog} using ${\rm Im}\det\bigl[m_u
m_u^\dag,\;m_d m_d^\dag\bigr] \equiv {\rm
Im}\det\bigl[S,\;S'\bigr]$. Jarlskog generalized to $n$
generations~\cite{Jarlskog87}, and found the invariant CPV measure
in terms of ``3 cycles", the trace of the cube of commutators of
quark mass squares, or ${\rm Im}\,{\rm tr\,} [S,\;S']^3$, which
looks considerably more complicated. To cut a long story short
(and somehow never invoked by Jarlskog in actual detail), note
that we are close to the $d$-$s$ (and $u$-$c$ as well) degeneracy
limit on the v.e.v. scale. In this degeneracy limit, the 4
generation world actually becomes the effectively 3 generation
world of 2-3-4 generation quarks\,! One sees now why
Eq.~\ref{eq:J234} would turn out to be by far the dominant, and
why $J$ in Eq.~\ref{eq:J3}, which could be written as $J(1,2,3)$,
is so tiny (the $10^{-10}$ gap!).

Out of the 3 independent phases in SM4, one is already measured in
$b\to d$ transitions, one could be emerging in a spectacular way
in $b\to s$ transitions.  A third subdominant phase can be
glimpsed from Figure~\ref{bdbsTri}. Since $V_{us}V_{ub}^*$ is
small, the resulting triangle by shrinking $\vert
V_{us}V_{ub}^*\vert\to 0$ is not much different from the
quadrangle. Thus, we have been a little cavalier in the notation
of $A_{234}^{sb}$, but again there is no doubt that
$J_{2,3,4}^{sb}$ of Eq.~\ref{eq:J234} is the predominant CPV
effect in SM4, and the relevant one for BAU. Judging from the
combined $10^{15}$ enhancement from $J$ to $J_{2,3,4}^{sb}$, it
seems sufficient to overcome the large gap of $10^{-10}$, even
taking into account the gauge factors that we have alluded to.

What about EWPhT? It is claimed that a first order transition is
not possible for SM4~\cite{FK08}. But perhaps strong Yukawa
couplings, beyond the unitarity limit of heavy $t'$ and $b'$
masses (perturbativity is lost), opens a new possibility, as EWSB
itself could be through the Nambu--Jona-Lasinio~\cite{Nambu}
mechanism with large Yukawa couplings.

\subsection{Tevatron/LHC Verification}

Given the developments at the Tevatron on $\sin2\Phi_{B_s}$ in the
past year~\cite{phisCDFtag,phisDzerotag,Tonelli,Paulini}, 2009
appears extremely interesting, while LHCb may not deliver physics
even by 2010. Judging from the recent performance of the Tevatron
accelerator and experiments, if the current central value
(consistent with Eq.~\ref{eq:sin2PhiBs}!) stays, we would have
evidence in 2009, perhaps even observation in 2010, \emph{if
Tevatron could continue running beyond 2009}. Regardless, once
LHCb has of order 0.5 fb$^{-1}$ data {\it analyzed}, whether one
has NP CPV enhancement or not, the whole situation would
precipitate.

But measurement of large $\sin2\Phi_{B_s}$, while exciting, does
not constitute proof for a 4th generation. The real litmus test,
as always, would be direct search. Current CDF limit gives
$m_{t^\prime} > 311$ GeV at 90\% C.L., using 2.8 fb$^{-1}$ data.
Again, once LHC data becomes available, the full terrain can be
covered in a straightforward way.

\subsection{Conclusion}

The gain of $10^{13}$ ($10^{15}$ if $m_{t',b'} \sim 600$ GeV is
used) in mass factors of Eq.~\ref{eq:J234} with 4 generations,
over Eq.~\ref{eq:J3} with only 3 generations, seem to give enough
CPV for generating BAU. \emph{Maybe there is a 4th Generation.}

In several years we should know whether the KM mechanism --- with
4th generations --- could provide sufficient CPV for BAU. It would
be amazing if what we find on Earth really has something to do
with (baryo-)Genesis\,!

\end{document}